\documentclass[12pt]{iopart}
\usepackage{iopams}  
\usepackage{graphicx}

\usepackage{hyperref}

\begin{document}

\title{Hot atomic vapors for nonlinear and quantum optics}

\author{Quentin Glorieux}

\address{Laboratoire Kastler Brossel, Sorbonne Universit\'{e}, ENS-Universit\'{e} PSL, Coll\`{e}ge de France, CNRS, 4 place Jussieu, 75252 Paris Cedex 05, France}
\ead{quentin.glorieux@lkb.upmc.fr}

\author{Tangui Aladjidi}

\address{Laboratoire Kastler Brossel, Sorbonne Universit\'{e}, ENS-Universit\'{e} PSL, Coll\`{e}ge de France, CNRS, 4 place Jussieu, 75252 Paris Cedex 05, France}

\author{Paul D Lett}
\address{National Institute of Standards and Technology, Quantum Measurement Division, MS 8424. Gaithersburg, MD 20882 USA, and}
\address{Joint Quantum Institute}
\address{National Institute of Standards and Technology and University of Maryland}
\address{College Park, MD 20742 USA}
\ead{paul.lett@nist.gov}

\author{Robin Kaiser}

\address{Université Côte d’Azur, CNRS, Institut de Physique de Nice, 06560 Valbonne, France}
\ead{robin.kaiser@inphyni.cnrs.fr}

\vspace{10pt}

\begin{abstract}
Nonlinear optics has been a very dynamic field of research with spectacular phenomena discovered mainly after the invention of lasers. The combination of high intensity fields with resonant systems has further enhanced the nonlinearity with specific additional effects related to the resonances. In this paper we review a  limited range of these effects which has been studied in the past decades using close-to-room-temperature atomic vapors as the nonlinear resonant medium. In particular we describe four-wave mixing (4WM) and  generation of nonclassical light in atomic vapors.  One-and two-mode squeezing as well as photon correlations are discussed. Furthermore, we present some applications for optical and quantum memories based on hot atomic vapors. Finally, we present results on the recently developed field of quantum fluids of light using hot atomic vapors.

\end{abstract}

%
%
\submitto{\NJP}

\section{Introduction}

Alkali atoms have long been used for studying light/atom interactions because their single valence electron simplifies their atomic level structure, making them easier to model theoretically. They also have strong transition strengths in the visible and infrared range.  For these reasons alkali atoms have been the focus of a large number of atomic physics experiments for many years, both in hot vapors and other systems. The strong transitions at optical wavelengths make the experiments more straightforward and the simplified level scheme makes the results easier to compare with theory.  Over the years the most affordable and available tunable lasers have changed from dye lasers to semiconductor diodes, and with that the favorite alkali has shifted from sodium to cesium and rubidium. 

Progress in certain classes of experiments has many times evolved from vapor cells, to atomic beams, to laser-cooled atoms, to Bose-Einstein condensates, and now, with advances in other supporting technologies, there is a trend towards a return to the simplicity of vapor cells for some experiments.  The development of vapor-cell atomic clocks and magnetometers \cite{Kitching}, as well as attempts to couple hot vapors to optical cavities \cite{Ritter} have pushed this trend along.

In this review paper, we will address the use of hot atomic vapors for non-linear optics, in particular for quantum optics with squeezing and quantum memories (section \ref{QO}) and for quantum fluids of light  (section \ref{QFL}). Further topics of non-linear optics using hot atomic vapors, such as electromagnetically induced transparency or Rydberg blockade will be addressed in separate review papers.

\section{Hot vapors for quantum optics}
\label{QO}

The alkalis have been studied over the years as effective 2-level systems that could be modeled quite completely, as optical pumping could prepare more complicated atoms into essentially 2-level systems.  As the experiments and theory progressed the field has moved beyond looking at  2-level systems.  The alkalis have the advantage that the atomic level structure is still simple enough that one can optically-pump the system into effective 3- or 4-level systems rather than the 2-level structure that quantum optics theory focused on in its early years.  One can, for instance, look at ground state coherences in simplified level structures.  Multi-level phenomena, such as electromagnetically-induced transparency (EIT) can be readily observed in alkali vapors.  The sub-natural-linewidth features that can be seen with EIT are evidence of quantum interference effects in these systems.  Similarly, other quantum effects, such as the generation of squeezed light, have evolved along a similar path.  The earliest squeezed light demonstrations relied on 2-level systems prepared in atomic beam experiments \cite{Slusher}.  More recent squeezed light demonstrations in atomic systems have relied on more complex level structures, which is what we will concentrate on here.

Squeezed light is essentially the reduction of noise in one variable of a system, at the cost of increased noise in a conjugate variable.  The amplitude and phase of a beam of light are tied together in the form of a Heisenberg uncertainty relation. This sets a lower limit on the product of the measurement noises in these two variables.  The square-root of this product is the “standard quantum limit” that represents the noise of the typically most “classical” state of light that we have: the coherent state.  The existence of this “limit” does not, however, prevent one from reducing the uncertainty in one of these variables, say the amplitude, at the cost of increasing the noise in the other (the phase).  Any variable that has a noise variance that is below the “standard quantum limit” is said to be squeezed. What is required to generate squeezing is a nonlinear process such as four-wave mixing that results in the correlated emission of photons. While there is nothing preventing us from generating squeezed light in a large number of situations, the simple fact that loss processes quickly reduce the state of the light back to the standard quantum limit, or worse, means that squeezed light is often tricky to generate and  difficult to work with.

\begin{figure}[htb]
    \begin{center}
	\includegraphics[width=.9\linewidth]{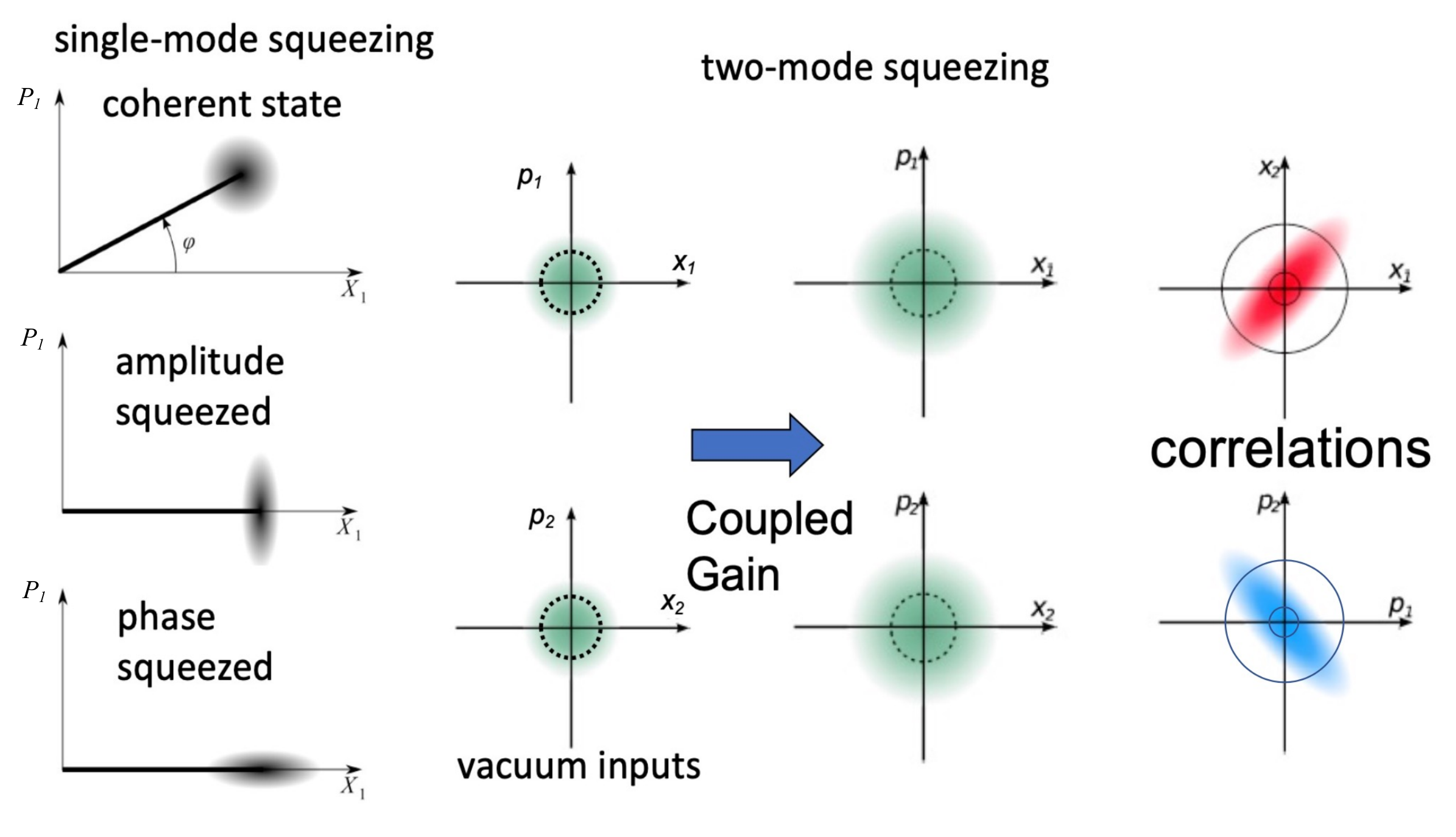}
    \end{center}
    \caption{Graphical representation of squeezed states of light.  (Left panel)  Single-mode squeezed state in phase space; a coherent state is a minimum-uncertainty state with equal uncertainties in all directions in phase space.  Amplitude- and phase-squeezed states are indicated, showing the reduced uncertainty in the respective properties of the mean-field vector.  (Right panel)  Two-mode squeezed state represented in a dual phase space (x1, p1) and (x2, p2).  In this case each input mode is a (minimum uncertainty) vacuum state.  After the nonlinear coupled gain process the amplified output states become noisier, thermal states of the field, but the correlations that they show on the diagonals in the (x1, x2) and (p1, p2) spaces are squeezed below the standard quantum limit.  Typically the intensity-difference and phase-sum of the fields are squeezed.}
    \label{fig:squeezing}
\end{figure}

The form of squeezing that connects the noise of, say, the amplitude and phase in a single mode of a field is called single-mode squeezing.  Single mode squeezing can also describe the noise correlations in generalized quadratures of the field; essentially linear combinations of amplitude and phase. Single mode squeezing is generated through the phase relationship of  pairs of photons emitted into the field mode.  Similar correlations can be generated through the emission of pairs of photons into separate field modes.  This “two-mode squeezing” creates, for example, intensity and phase correlations between two beams of light.  In the most common case the amplitude or intensity-difference and the phase-sum of two beams can be squeezed, creating entanglement in these variables. The individual beams are typically quite noisy in these circumstances, but they are correlated to below the shot noise level.  Once again, generalized quadrature combinations of the two field modes can also be squeezed.  This relationship is pictured in Fig.~\ref{fig:squeezing}.

\begin{figure}[htb]
    \begin{center}	\includegraphics[width=.8\linewidth]{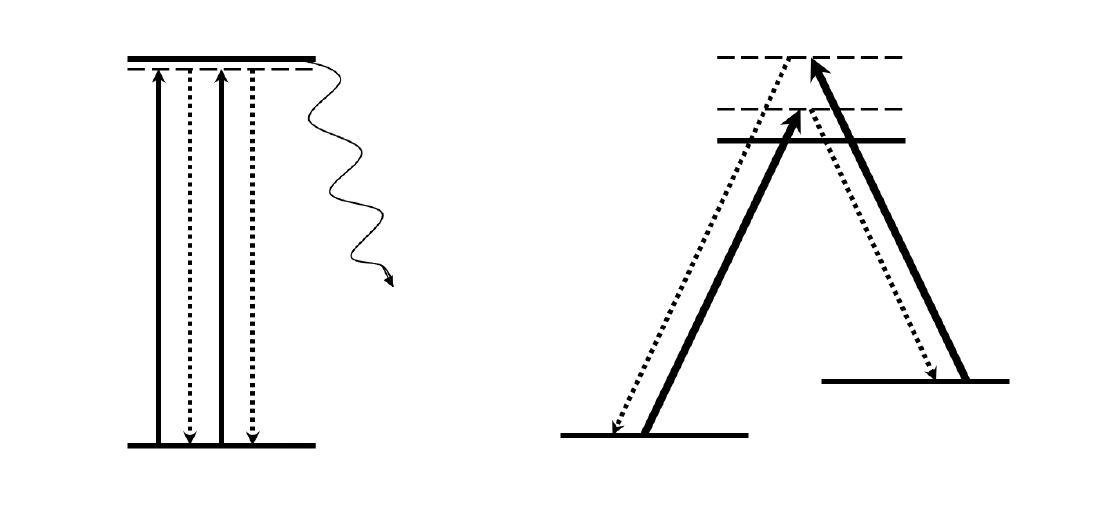}
    \end{center}
    \caption{Sketches of the atomic level diagram.  (Left panel)  Degenerate 4-wave mixing in a 2-level (ground-excited) system.  The transition is driven near resonance, and spontaneous emission from the upper level adds noise to the system.   (Right panel). Off-resonant 4-wave mixing in a 4-level system.  The two upper states are virtual states, significantly detuned from the atomic transition.  The coherence is driven between the two ground states, and the upper states are off-resonant and there is no significant excitation of these states.}
    \label{fig:levels}
\end{figure}

Photon correlation measurements can often be conditional measurements, where losses cause the measurements to be inefficient, but do not obscure the effects of quantum correlation. The observation of squeezing, on the other hand, requires very low loss.  Usually this is seen as a problem for hot vapors, as light near the optical transition frequencies (where interactions are strong) will tend to be absorbed to some degree by the Doppler-broadened vapor.  The 2-level system also suffers from spontaneous emission noise, as indicated in Fig.~\ref{fig:levels}. Since real absorption transitions to the upper level are likely, there is always the possibility that uncorrelated spontaneous emission will appear in the detected light as well. This emission noise, plus the noise due to the re-absorption of the light emitted in the nonlinear process, degrades the measured squeezing.  The 3- or 4-level system (depending on whether you count virtual or only real levels – you would need 4 levels to allow the transitions with proper polarizations) relying on ground-state coherences does not require any of the light to be very close to a real absorption resonance, and so can be operated essentially without spontaneous emission from the excited state(s)  \cite{glorieux2010double}.

\begin{figure}[htb]
    \begin{center}
	\includegraphics[width=.9\linewidth]{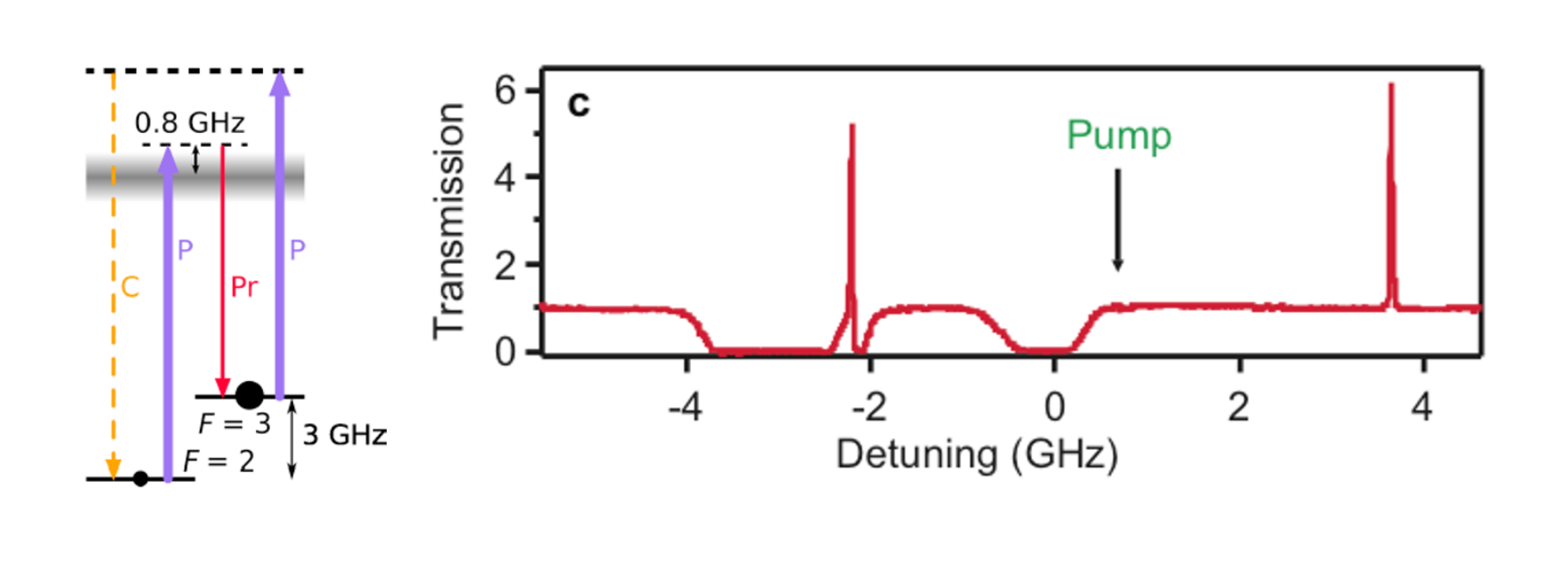}
    \end{center}
    \caption{Atomic level diagram for $^{85}$Rb.  The two ground states are separated by approximately 3 GHz.  The transmission spectrum (gain $>$ 1, and absorption $<$ 1) of the experiment (shown in Fig. 5), where a pump beam is crossed by probe beam which is scanned in frequency and detected at the output.  The Doppler-broadened absorption features represent transitions from the two ground states to the excited state, while the narrow gain peaks represent the 4-wave mixing gain, and are located approximately $\pm$3 GHz from the pump.}
    \label{fig:gain}
\end{figure}

The reliance of the 3- or 4-level 4WM scheme on off-resonant excitation and ground-state coherences, rather than the excited state coherence of a 2-level system allows it to get around the loss as well as the spontaneous emission.  As can be seen in Fig.~\ref{fig:gain}. for $^{85}$Rb 4WM, the pump can be located just at the edge of the Doppler-broadened absorption profile, while the 4WM probe and conjugate frequencies are either also just on the edge of the absorption profile, or well away from it.  The gain is reduced by operating with the pump $\approx$ 1 GHz off resonance, but with the loss being low ($\approx$ 10 $\%$ or less on the closer-to-resonance gain feature and much less on the far-from-resonance gain feature), a single-pass gain of $\approx$ 5 results in strong measured squeezing.  (Without loss a gain of 5 would result in more than 9 dB of squeezing.)
 The situation is particularly favorable for Rb and Cs, where the Doppler-broadened absorption line can be avoided almost completely by the light generated in the 4WM process. Even in the case of K, however, where the ground state hyperfine splitting is smaller than the Doppler width of the transition and leads to substantial loss, significant 2-mode squeezing can be measured \cite{Swaim}.

\subsection{Photon counting versus continuous detection}
Suffice it to say that there are many ways to obtain and test for quantum correlations.  While this review will concentrate on the generation of continuous-wave squeezed light from hot vapors, many groups have generated non-classical photon-counting correlations as well, and a brief mention of some of this work is included here.  A common non-classical feature that is demonstrated for photon-pair generation is a violation of the Cauchy-Schwarz inequality, relating the cross correlations and auto-correlations of pairs of photons.  While violating this inequality (that the square of the cross correlation of two beams should be less than or equal to the product of the autocorrelations of the two beams) does demonstrate that the light is non-classical, it is something of a binary test.  That is, violating the inequality says that the fields are non-classical.  Violating it by a larger amount says nothing more than that.  Typically the magnitude of the violation is larger if the average intensities are small.  Demonstrating a violation in the squeezed-light intensity regime is challenging, but has been done \cite{Marino}.  

While generating correlated photons in a hot vapor is not difficult in principle, getting the correlated photons out of the vapor without reabsorption is often a significant barrier to demonstrating the non-classicality of the field.  Nonetheless there have been a fair number demonstrations of this sort. Most of these experiments have been performed in Rb vapor; presumably this is mostly because of the availability of lasers in the desired range. The various demonstrations include using backwards non-degenerate 4WM in a double-lambda scheme \cite{Chen} \cite{Podhora}, and using a diamond-type or ladder schemes \cite{Willis}\cite{Ding}\cite{LeeY}.
The absorption in the vapor can also be reduced by optimizing an electromagnetically-induced transparency window \cite{Zhu}. 

One can also use these photon correlations to produce a heralded source of single photons.  The Fock state produced in this way is also, of course, highly non-classical.  Such a source was constructed using a 4WM process, also in Rb vapor \cite{MacRae}.

\subsection{Squeezing }

\subsubsection{Polarization self-rotation}
An exceptionally straightforward technique for generating squeezed light in a hot alkali vapor is that of polarization self-rotation or cross-phase-modulation squeezing.  This technique was first proposed in \cite{Matsko} and observed in \cite{Ries}.  The technique has been further developed in recent years and has proven to be versatile.  

Polarization self-rotation describes the rotation of the polarization of elliptically polarized light as it propagates through a medium. The interaction of an initially isotropic Kerr medium with an elliptically polarized light field causes the medium to become circularly birefringent: the two circular components of different intensities will propagate with different phase velocities.  In near-resonant atomic vapors the self-rotation can be viewed as being due to optical pumping and unbalanced ac-Stark shifts in the medium caused by unequal intensities of circularly-polarized components of an input light field.  For a linearly polarized field there would be no polarization rotation in this picture but the same interactions result in a coupling of the two circularly-polarized components of the vacuum field.  A linearly-polarized field can be analyzed as two circularly-polarized components and thus the interaction produces a cross-phase modulation effect between the pump field and an orthogonally-polarized vacuum field, which becomes squeezed.  Such self-rotation is present in any Kerr medium, but atomic vapors are particularly favorable because of the strong nonlinearities near resonance.  

\begin{figure}[htb]
    \begin{center}
	\includegraphics[width=.8\linewidth]{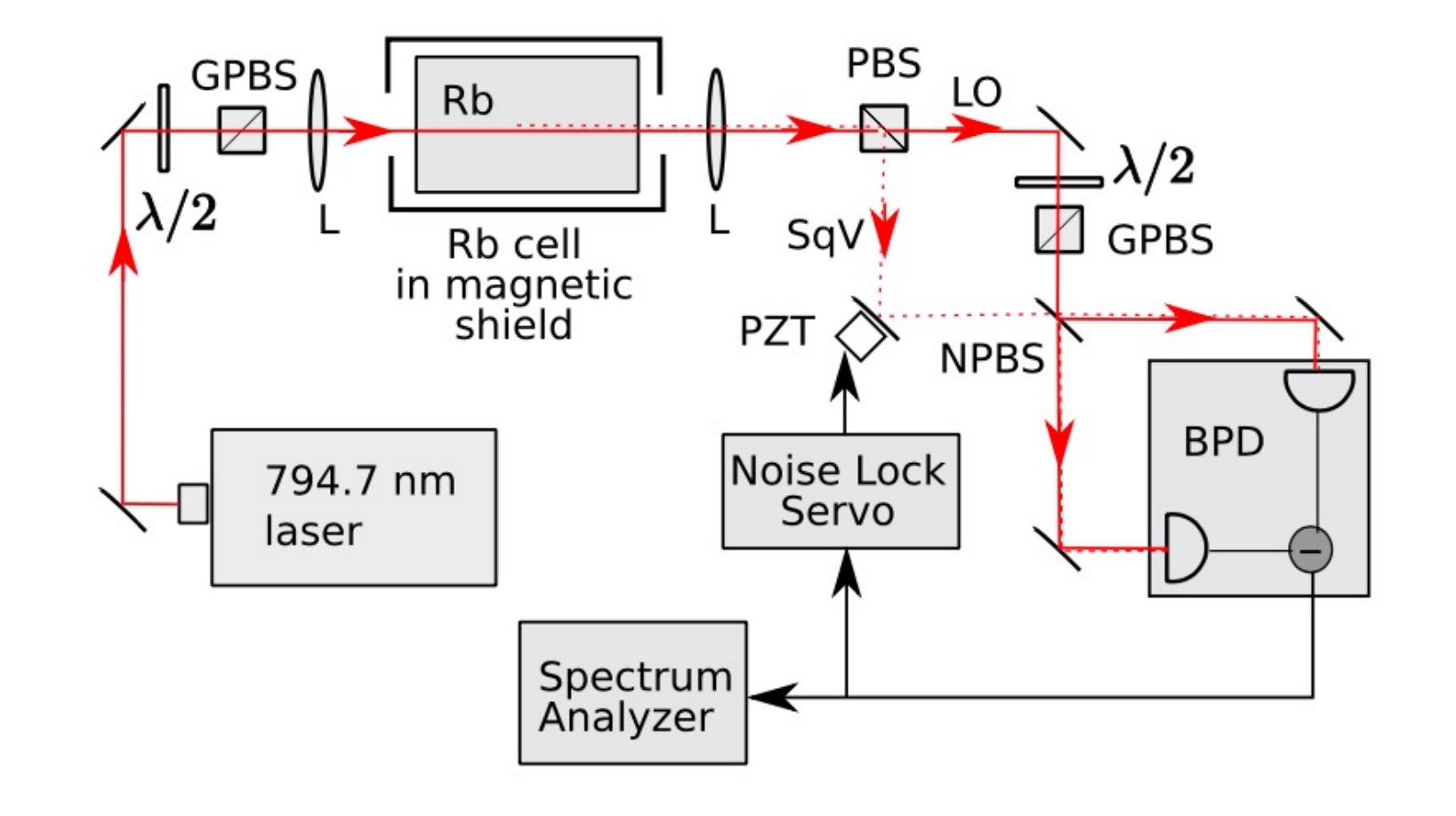}
    \end{center}
    \caption{Figure of a polarization self-rotation set-up.  SqV is the squeezed vacuum field.  LO is the local oscillator.  NPBS is a non-polarizing beamsplitter and GPBS is a Glan polarizing beamsplitter. BPD is a balanced photodiode detector.  PZT is a piezo-electric translation actuator on a mirror. In the experiment there is actually a $\lambda$/4 waveplate (not shown here) just before the cell to introduce some ellipticity and thus obtain squeezing after. 
 (From Ref.  \cite{Mikhailov2}.)}
    \label{fig:selfrot}
\end{figure}

The laser power requirements are modest – a few milliwatts to tens of milliwatts tuned near an atomic resonance.  This and a rubidium vapor cell a few cm long are all that are, in principle, required to generate single-mode squeezing.  As usual, the detection apparatus is somewhat more complicated, but not tremendously so, leading to its conceptual appeal.  This system is elegant in that the original pump beam can play the role of the local oscillator phase reference for detection, although it may need to be attenuated first.  A wave plate rotates the polarization at the output from the vapor cell to allow the beams to be separated by a polarizing beamsplitter. Using the pump as a local oscillator removes some of the sensitivity of homodyne detection to beam distortions going through the system.  The signal and pump (local oscillator) can be recombined onto a balanced detector for homodyning (after phase-shifting and possible attenuation of the LO beam), as indicated in Fig.~\ref{fig:selfrot}.

Single-mode squeezing at low frequencies has been observed using this technique  – down to approximately 100 Hz \cite{Horrom}\cite{Mikhailov1}. The range of squeezing could extend to 10 MHz or 20 MHz \cite{Ries}.

One of the drawbacks of the technique is that the demonstrated squeezing has so far been somewhat limited in magnitude, with about 3 dB being the present limit \cite{Barreiro}.  The squeezed light generation is sensitive to magnetic fields, and 3-layer magnetic shielding is often used.  It is also susceptible to  re-absorption loss in the vapor, as it is operated near-resonance.  The Rb cell is sometimes filled with a few hundred Pa (1~Torr $\approx$ 133.32~Pa) of Ne buffer gas to assist in keeping the atoms which are optically pumped confined to the interaction region, so as to reduce the loss \cite{Mikhailov2}.

\subsubsection{Four-wave mixing based on ground state coherences}

\begin{figure}[htb]
    \begin{center}
	\includegraphics[width=.9\linewidth]{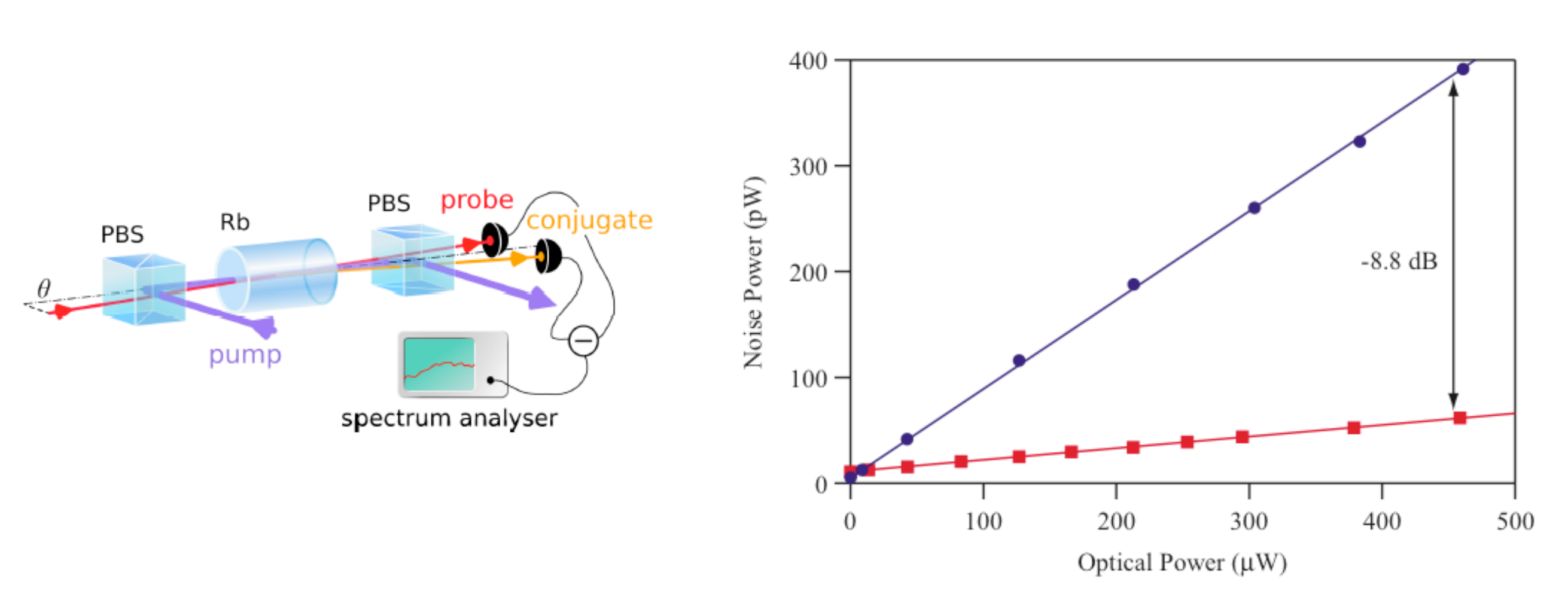}
    \end{center}
    \caption{Sketch of experiment for generating 2-mode squeezing from 4-wave mixing.  PBS is a polarizing beamsplitter.  A pump is sent through a Rb vapor cell and a probe beam, orthogonally polarized to the pump, is intersected with the pump at a small angle.  A conjugate beam is generated at the symmetric angle on the other side of the pump. The probe and conjugate beams are detected, subtracted, and the noise power is recorded on an rf spectrum analyzer.  The plot on the right shows the noise power at 1 MHz (in a 300 kHz bandwidth) as a function of the total optical power detected.  The black points represent the shot noise level and the red points show the noise from the 2-mode intensity-difference squeezing, showing about 8.8 dB less noise at high powers.}
    \label{fig:2mode}
\end{figure}

\paragraph{Two-mode squeezing based on double-lambda systems.}
One is able to generate both 1- and 2-mode squeezing in atomic vapors using double-lambda 4-wave mixing processes \cite{glorieux2010double}.  While the description of single-mode squeezing itself is more straightforward, the actual generation of two-mode squeezing is, at least conceptually, extremely simple, so we will start with describing the generation of 2-mode squeezing in $^{85}$Rb.  In principle at least, such squeezing can be generated by simply sending an appropriately-tuned and sufficiently-intense pump laser through a warm Rb vapor cell (see Fig.~\ref{fig:2mode}).  Of course, there are more details to be considered.

A level scheme as indicated on the right in Fig.~\ref{fig:levels} can be used to describe the generation of correlated “twin beams” of light. A pump beam, detuned about 1 GHz from the atomic D1 line (near the edge of the Doppler-broadened absorption profile, as in Fig.~\ref{fig:gain}), is introduced into an approximately 1 cm long Rb vapor cell, held at about 120 $^{\circ}$C. This gives sufficient vapor density for good 4WM gain, and allows for some mismatch in the phase matching.  (Longer cells put tighter constraints on the phase-matching angle.)  About 0.5 W of cw laser power is typically used in an approximately 800 $\mu$m diameter beam. (It is the intensity that is important for the efficiency of the process, not the power, but this allows for a reasonable-sized central gain region, with approximately constant gain, that is easy to send a probe beam through.) As the $\chi^{(3)}$ nonlinearity of the vapor is strong, the intensity-dependent index of refraction for the various beams becomes important for phase-matching, and exactly co-linear phase matching is not favored.  The correlated “probe” and “conjugate” beams, in the language of 4-wave mixing, have their highest gain at a small angle (about 0.5 degrees) on either side of the pump.  The probe and conjugate beams are also polarized orthogonally to the pump beam.

The angle between the beams and the fact that they are orthogonal in polarization to the pump allow the twin beams to be easily separated from the pump.  This is convenient because these beams are only about 3 GHz away from the pump in frequency, and so it is difficult to efficiently filter-out the pump light without using carefully constructed cavities or using homodyne detection techniques. Because of the strong optical pumping, no magnetic shielding is needed around the cell.  The above conditions can provide a single-pass gain on an injected probe beam in the range of 5 to 10, so that multi-spatial-mode imaging through the gain medium is possible with this set-up without a build-up cavity that would spatially-filter an image.  

The 4WM process takes place with little re-absorption in the vapor.  While near-resonance 4WM schemes have demonstrated much larger gain, it is the lack of reabsorbtion loss here that is key to observing strong quantum correlations.  Using an injected probe beam intensity-difference squeezing of approximately 10 dB has been demonstrated through direct intensity detection by this method \cite{glorieux2011quantum}. The noise on the probe seed should be near the shot noise limit for this to work.  Homodyne measurements have demonstrated approximately 5 dB of quadrature squeezing and entanglement (simultaneous amplitude-difference and phase-sum squeezing).  The difference is due to the increased cost of local-oscillator misalignment as the squeezing level is increased (increasing the anti-squeezing and the noise cost of measuring uncorrelated spatial modes surrounding the desired ones \cite{Gupta}).

This type of two-mode squeezing has been demonstrated in both $^{85}$Rb and $^{87}$Rb, on both the D1 and D2 lines \cite{Pooser1}.  It has also been demonstrated in Cs \cite{Ma1} \cite{Ma2}, and K \cite{Swaim}. As mentioned above, the basic requirement is that the Doppler-broadened absorption profile in the vapor not overlap the frequencies of the emitted probe and conjugate beams.  This is not completely true, inasmuch as the situation in potassium begins to violate this statement; with a fine structure splitting of only 460 MHz and a Doppler width of $\approx$ 850 MHz, the probe beam frequency suffers large absorption, but measurable squeezing ($>1$ dB) can still be obtained.
An interesting extension of this approach is the generation of pulsed relative-intensity squeezing, by pulsing the pump beam in the 50 nanosecond range \cite{agha2011time}.

While bright twin beams can be created by seeding the process along a particular direction, there is no physical restriction on the 4WM process that limits it in this way.  Vacuum-seeded 4WM takes place on an entire cone centered on the phase-matching angle around the pump beam.  This azimuthal freedom, combined with the phase-mismatch that is allowed, results in a large number of spatial modes that can be squeezed.  Thus, without the restrictions of a cavity, multiple spatial modes, in the form of “twin images” can be generated \cite{Boyer}.  The gain bandwidth of the process is not very large – approximately 20 MHz for typical parameters in the case of Rb, set by the power-broadened natural linewidth \cite{Liu}.
Moreover, with a pulsed homodyne detection scheme, pulsed bipartite entanglement has been measured \cite{glorieux2012generation}.

\paragraph{Single-mode squeezing with double-lambda systems.}

Once we can see how twin-beams or 2-mode squeezing is generated in this process, it is easy to imagine swapping the roles of the pump and the probe and conjugate beams.  We change the intensities of the beams so that in this case we introduce two relatively-strong pump beams along what were the probe and conjugate paths, at frequencies 6 GHz apart in Rb, as is the case discussed above. We can then see that a beam injected between them, along what was the pump path in the 2-mode case, and with a frequency half-way between the pumps, will experience phase-sensitive amplification through a similar 4WM process.  It turns out that the overall detuning of these three beams needs to be shifted somewhat from that used in the 2-mode squeezing experiments, but when this is done single-mode squeezing can be obtained \cite{Corzo}.  

This 4WM scheme is phase-sensitive and requires the locking of the phases of the two pump beams, as well as the probe if a bright probe beam is introduced.  For the correct choice of the relative phase ($2*\theta_{probe} - \theta_{pump1} - \theta_{pump2}  = 0 $) the 4WM process then produces stimulated absorption of the probe beam.  If the bright injected probe beam is blocked, and the process is vacuum-seeded, the result is single-mode vacuum squeezing.  As in the 2-mode squeezing process, the phase matching requirements are sufficiently loose that a number of independent spatial modes can be squeezed in this configuration.  This technique of generating single-mode squeezing is somewhat limited in its ability to generate strong squeezing in this geometry.  The generation of 2-mode squeezing based on each one of the two pump beams can lead to “parasitic” 4WM processes that simply add noise to the desired process.  This limits the useful gain of the process such that approximately 3 dB to 4 dB of single-mode squeezing can be obtained before the parasitic processes start to degrade the measurements.  The bandwidth of the process is similar ($\approx$20 MHz) to that of the 2-mode squeezing process.

If the relative phase between the injected probe beam and the pump beams is shifted by $\pi/2$ the 4WM process results in phase sensitive gain instead of absorption.  The advantage of this configuration is that, in the limit of large gain the amplification is noiseless.  With the correct choice of phases, for example, if one sends in a probe with a given signal-to-noise ratio (SNR) on an encoded intensity modulation, the output can be amplified and  retain the same SNR.  The more common phase-insensitive gain always comes with a 3 dB “noise penalty” – that is, the SNR out will be reduced by a factor of two (again for large gain).  While theoretical arguments can be given that show that this must be so, the physical reason in the present case can be seen in the example of the 2-mode squeezing discussed above.  In that case one can view the 4WM process as a phase-insensitive amplifier for a bright probe beam that is introduced into one input port.  While this beam is amplified through the 4WM process, the conjugate input port is simultaneously being fed by a vacuum input, which is also amplified, and these two outputs are added.  Thus, there will always be additional noise added to the signal in this situation.  The price paid to obtain noiseless gain is the phase sensitivity of the gain.

One way in which phase sensitive gain can be used is to create “perfect detectors” that can compensate for detector inefficiency.  Homodyne detection is already phase-sensitive, so one can insert an appropriate phase-sensitive gain element to amplify the signal before an inefficient photodiode.  Again, in the limit of large gain, this amplification can compensate for detector inefficiency, even for signals of a quantum nature.  As most detectors have relatively good (say 50 $\%$) efficiency, even modest gain can have a significant effect, as demonstrated in  \cite{Li}.

Additional geometries can be used to enable the generation of frequency-degenerate twin beams \cite{Jia}, which could be combined with the correct phase to generate bright single-mode squeezing.  In this case the geometry that is used is essentially the same as that used for generating single-mode squeezing with two pumps discussed above. In this case, however, the phase matching is out-of-plane, so that photon pairs are generated in twin beams instead of directly into a single beam.  One advantage of this configuration is that stronger squeezing can be obtained as fewer parasitic 4WM noise sources are excited.

In addition, given the relatively small isotope shift in Rb the squeezing can be generated on-resonance for $^{87}$Rb atoms through a 4WM process in $^{85}$Rb \cite{Kim}. In this case bright non-degenerate two-mode squeezed states are generated near the D1 line in hot $^{85}$Rb. The authors were able to generate two-mode squeezed states where the twin beams were simultaneously on resonance with the F = 2 to F' = 2 transition and the F = 1 to F' = 1 transition in $^{87}$Rb. 
An intensity-difference squeezing of 3.5 dB was obtained.  These authors again used this approach for single-mode squeezed light, using 4WM in $^{85}$Rb, to produce squeezing on-resonance for $^{87}$Rb. In this case they measured fluctuations in one of the twin beams and used a feedforward scheme to reduce the fluctuations in the twin.  About 2 dB of squeezing was produced near the atomic resonance \cite{feedforward}.

\subsection{Metastable gases}
While not exactly a “hot” vapor, metastable helium (He*) is an interesting alternative to alkalis with regards to phenomena like EIT and squeezed light generation.  While not used as commonly as alkali vapors, the level structure, starting from the metastable state 19.8 eV above the ground state, is sufficiently simple that laser cooling \cite{Aspect88, Kaiser91, Lawall} and EIT \cite{Goldfarb} are possible. A 4WM gain process \cite{Neveu1} and even squeezed light generation \cite{Neveu2} using such atoms have been demonstrated in this system. The metastable $2^3S_1$ state has a lifetime of approximately 8000 s, and is typically excited in an rf discharge.  Penning ionization is a potential problem in such a highly excited system (colliding metastable atoms de-exciting through ionization of one of the pair), but optical pumping to states that are not allowed to scatter can reduce this issue greatly \cite{Shlyapnikov}. Gas densities in this system are such that the atoms under study are generally confined to the exciting laser beams by diffusive transport in background He instead of ballistically leaving the excitation region.

Neveu, et al. \cite{Neveu2} demonstrated the generation of squeezed vacuum states of light by four-wave-mixing-enabled coherent population trapping in a metastable helium cell at room temperature. 
Unlike the off-resonant alkali systems discussed above, they work on-resonance with an atomic transition. Helium atoms are pumped into their metastable state $2^3S_1$ by means of an electrical discharge at 80 MHz. This results in a density of He* of about $2 \times 10^{11}$ cm$^{-3}$. 
A fiber laser at 1.083 $\mu$m, resonant with the D1 transition of He*, is used.  Linearly polarized pump and probe beams interact with atomic sublevels of the $2^3P_1$ and the nearby $2^3P_2$ states, respectively.  A 55 mW pump beam results in strong EIT and a transparency window that reduces the absorption losses, as well as a slow group velocity that enhances the interaction and allows the observation of squeezing in spite of the large optical depth (4.5) of the vapor.  The system is unusual in that it is on-resonance, and therefore the squeezing is limited to frequencies within the transparency window resulting from coherent population trapping ($\approx$ 800 kHz).

Single-mode squeezing is observed on a vacuum-seeded probe beam that is degenerate in direction and frequency with the pump, but in the orthogonal polarization. The minimum noise observed corresponds to 0.51 dB of squeezing. This is limited by losses from residual absorption on the D2 transition and to the associated spontaneous emission, as accurately modeled by the authors.

\subsection{Optical and Quantum memories}
In this section, we give an overview of hot atomic vapors for optical and quantum memories.
A quantum memory for light is a device that can efficiently delay or store the quantum states of light fields.
Since the seminal paper of Duan-Lukin-Cirac and Zoller (DLCZ) \cite{duan2001long} considering quantum memories in atomic ensembles for long distance quantum communication, hot atomic vapors have been a medium of choice for storing and retrieving light.
With the emergence of the concept of dark state polaritons \cite{fleischhauer2002quantum}, electromagnetically-induced-transparency (EIT) has initially emerged as the main protocol for atom-based memories \cite{liu2001observation}. 
Early works demonstrating the coherent storage of light \cite{phillips2001storage} and the storage of quantum states \cite{van2003atomic,julsgaard2004experimental} have all been obtained in hot atomic vapors.
Excellent reviews cover this field extensively \cite{RevModPhys.75.457,lvovsky2009optical}, and therefore we take a different perspective here.
We focus on various alternative protocols and highlight the importance of multiple modes in optical quantum memories.

\subsubsection{Gradient-Echo-Memory}
Aside from EIT protocols and Raman coupling, another way of storing quantum information relies on photon echoes \cite{PhysRevLett.13.567}.
In atomic vapors, photon echoes were initially proposed for classical optical storage  \cite{carlson1983storage}.
However, by taking advantage of a Doppler-broadened transition, it is possible to store and reconstruct completely the quantum state of a single-photon wave packet via a protocol known as controlled reversible inhomogeneous broadening (CRIB) \cite{moiseev2001complete}.
Initially implemented in solid-state systems \cite{PhysRevLett.100.023601}, this approach has been very successful in hot atomic vapors especially for storing and retrieving multiple spatial modes \cite{nunn2008multimode}.
The technique is known under the name of gradient echo memory (GEM) and can be understood as follows.

Consider an assembly of two-level atoms, and imagine you have a physical mechanism to modify the transition frequency (say linearly) as function of the distance along $z$.
Then, when a near-resonant light pulse propagates through the atomic cloud (along $z$), the various frequency components of the pulse are absorbed preferentially at different positions $z$ as they become more resonant with the frequency shifted transition.
The quantum state of the electric field $E$ is then transferred to the atomic polarization $\alpha$.
The atomic polarization will subsequently dephase, since we have modified the transition frequency along $z$. 
The idea behind the gradient echo memory is to flip this frequency gradient after a given time $\tau$, such that the atomic polarization will perfectly rephase at time $2\tau$.
The atomic polarization can then be converted back to the field and the information is released from the memory.

\begin{figure}[htb]
    \begin{center}
	\includegraphics[width=.5\linewidth]{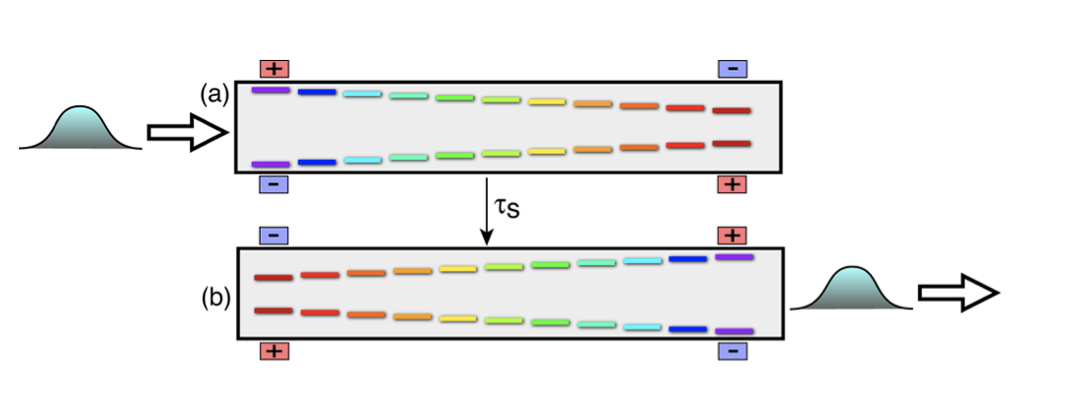}
    \end{center}
    \caption{Sketch of Gradient echo memory protocol. Figure adapted from \cite{PhysRevLett.101.203601}}
    \label{fig:gem}
\end{figure}

From a formal perspective, this protocol is summarized by two coupled differential equations:
\begin{eqnarray}
    \frac{\partial}{\partial t} \alpha(z,t) = - [ i \eta (t) z] \alpha(z,t) + i g E(z,t),\\
    \frac{\partial}{\partial t} E(z,t) = i \mathcal{N}\alpha(z,t),
\end{eqnarray}
where $\mathcal{N}$ is  the effective linear density of atoms and $g$ is the atomic coupling strength.
The spatially varying frequency shift (taken as linear here) is given by $\eta(t) z$ where the slope $\eta(t)$ can be flipped in time.
Within this formalism, the efficiency $\sigma$ of this protocol has been derived as $\sqrt{\sigma}=1-\textrm{e}^{-2\pi g \mathcal{N}/\eta}$.
Therefore to increase the memory bandwidth $\eta$, it is mandatory to increase the atomic density to keep the efficiency constant.
This potentially ultra-high efficiency is due to the absence of re-absorbtion of the light pulse after rephasing (as opposite to EIT protocols). 
Indeed, with the reversed frequency shift, the pulse is not resonant with the atomic medium anymore and no re-absorption occurs.

Experimental implementations in hot atomic vapors follow slightly modified protocols in order to take advantage of the long-lived ground-state coherence in three-level systems.
With three-level atoms in a $\Lambda$ configuration, instead of storing the information in the ground-excited coherence, a Raman coupling scheme is used to transfer the information into the ground-ground coherence.
It relies on the use of an intense control laser field near two-photon resonance with the quantum state one might want to store in the memory.
To shift linearly the resonance frequency of this two-photon transition, it is possible to apply a magnetic field gradient along the propagation direction.
It lifts the degeneracy of atomic Zeeman sub-levels and induces a controlled broadening \cite{hetet2008photon}.
Similar to the two-level configuration, flipping this magnetic field gradient will induce a rephasing of the ground state coherence and allows for the light to be retrieved from the memory.

There are actually several advantages of using this three-level scenario instead of the simpler two-level one.
First, the long-lived ground state coherence allows for a long storage time with high efficiency \cite{hosseini2011high}.
Second, the Raman coupling beam can be turned off during the storage time reducing spontaneous emission and improving noise properties to the quantum limit \cite{hosseini2011unconditional}.
Finally, coherent manipulation of optical pulses has been achieved using this approach.
Because, no light can be re-emitted from the memory if the Raman coupling beam is off, it has been shown that several pulses could be stored and retrieved as first-in-first-out (FIFO) or first-in-last-out (FILO).
Indeed, when two pulses (pulse A followed by pulse B) are stored in the memory, if the gradient is reversed after a time $\tau$, in the presence of the Raman coupling the two pulses are re-emitted after a time $2 \tau$ (pulse B followed by pulse A) in FILO configuration.
However, if the Raman coupling beam is off at time $2\tau$ the coherence will start to dephase again (in the opposite direction). A subsequent inversion of the gradient at time $\tau'$ will induce a second rephasing at time $2(\tau+\tau')$. 
If the Raman beam is present, the two pulses will exit (pulse A followed by pulse B) in a FIFO configuration \cite{hosseini2009coherent}.

The second important aspect of Gradient Echo Memory in hot atomic vapor is the possibility to store (and manipulate) multiple spatial modes in the transverse plane of the field \cite{PhysRevLett.109.133601}.
For example images have been stored and selectively erased using a third laser to destroy the coherence locally \cite{clark2013spatially}. 
It was also possible to combine temporal multiplexing with multiple spatial modes and store a short movie of 2 frames in a hot atomic vapor \cite{glorieux2012temporally}. 
Storing multimode transverse images in atomic media is crucial in constructing large-scale quantum networks but a major limitation on the storage time in these multi-mode systems is a result of atomic diffusion \cite{PhysRevLett.100.123903,shuker2008storing}.
However, an interesting direction has been opened up recently by combining the use of a genetic algorithm with phase-shift lithography to construct the optimal phase for an arbitrary transverse image. This can drastically diminish the effect of diffusion during the storage in warm atomic vapor \cite{chen2021genetic}.

\subsubsection{Spatial entanglement}
With the rapid development of spatially resolving single-photon detectors, spatially structured multidimensional entangled states start to play a key role in modern quantum science.
For example, spatial entanglement is known to enhance the capacity of quantum memories based on DLCZ-like protocols including GEM.
Experimentally this hybrid entangled state has been prepared via a two-mode squeezing Raman interaction, which creates pairs of Stokes photons and spin-wave excitations with anti-correlated momenta \cite{dkabrowski2017einstein}.
The medium used was rubidium-87 vapor with krypton buffer gas at 0.01 Pa ($\approx$ 1 Torr) to make the atomic motion diffusive. 
Interestingly, this approach is complementary with previous works, where time-delayed EPR correlations were demonstrated  \cite{marino2009tunable} but it appears to be more robust since the storage time in a quantum memory has been improved by two orders of magnitude and the violation of the EPR inequality was stronger than previously reported \cite{Marino}.

\subsubsection{Coherent population oscillation}
An interesting alternative to coherence-based atomic memories is to use a protocol which involves only the atomic populations.
In this case, the storage would become immune to decoherence effects described previously and  longer storage time could be achieved.
Coherent population oscillation (CPO) is a coherent phenomena which occurs in a two-level system when two detuned coherent electric fields of different amplitudes drive the same transition.
When the detuning between the fields is smaller than the decay rate of the upper level, the dynamics of the saturation opens a transparency window in the absorption profile of the weak field \cite{neveu2017coherent}.
Similar to GEM, CPO can be improved using a three-level $\Lambda$-system where two CPOs may occur in opposite phase on the two transitions, leading to a global CPO between the two lower states \cite{eilam2010spatial}.
Experimental demonstrations of light storage were performed using metastable helium (He*) vapor at room temperature \cite{maynard2014light}, as well as with warm cesium \cite{PhysRevA.90.043803}.
Recent theoretical works have also demonstrated the preservation of the pulse's temporal shape during storage using CPO \cite{PhysRevA.98.013808} and the possibility to extend this scheme to the storage of a quantum field \cite{PhysRevA.100.013820}.

\subsubsection{Novel protocols for ultra-long storage time}
Novel approaches have emerged recently for drastically improving the storage time in hot atomic vapors.
As we have seen, in dense gases, atomic collisions usually dominate the lifetime of the spin coherence and therefore limit the storage time to a few milliseconds.
It is therefore highly beneficial to use a storage scheme that is insensitive to spin-exchange collisions.

\begin{figure}[htb]
    \begin{center}
	\includegraphics[width=.5\linewidth]{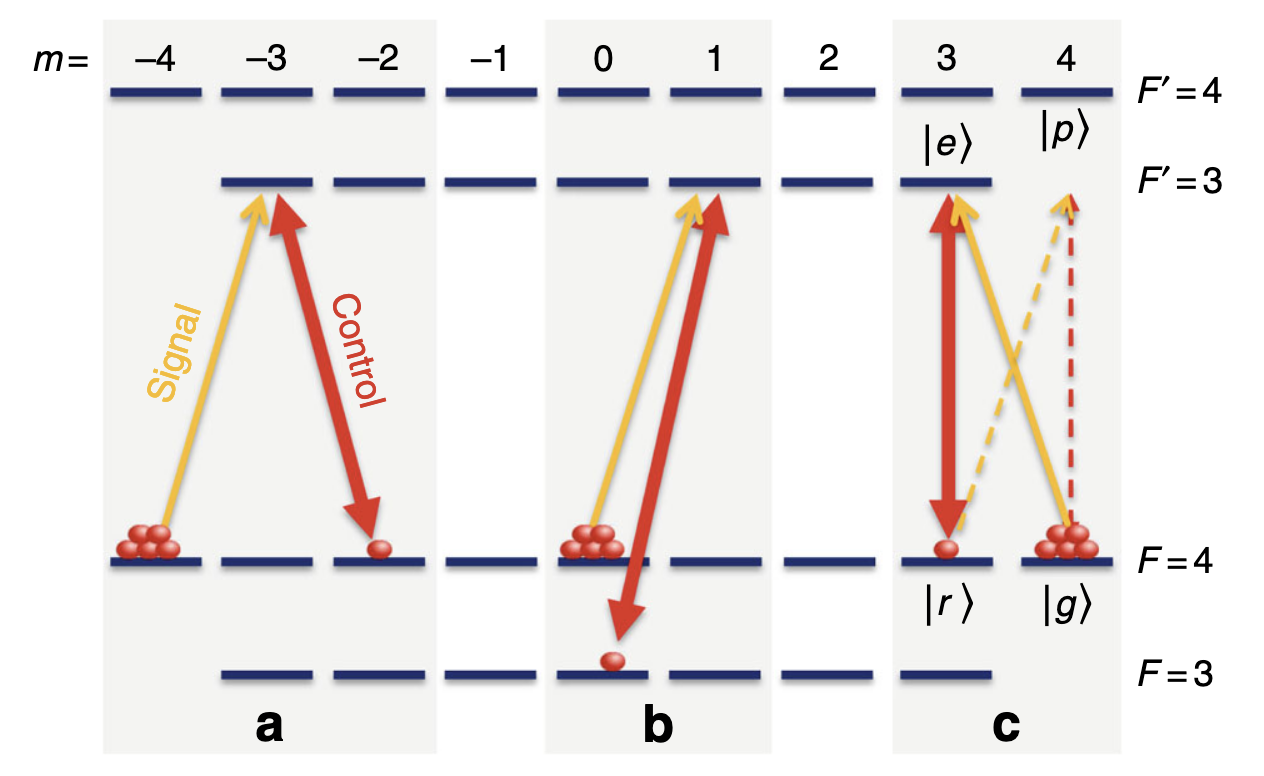}
    \end{center}
    \caption{Three configurations for light storage in caesium ground levels. a) Zeeman coherence with $\Delta m =2$, b) hyperfine coherence with $\Delta m =0$, c) Zeeman coherence with $\Delta m =1$ associated with the spin orientation moment, known to be unaffected by spin-exchange collisions. Figure taken from \cite{katz2018light}.}
    \label{fig:spin}
\end{figure}

The key point of this approach is to take advantage of the nuclear spin orientation instead of the electron spin. 
The multiple spin states are characterized by the hyperfine spin $F$ and its projection $m$ onto the quantization axis.
Commonly used light storage techniques are based either the Zeeman coherence $\Delta m =2$ (Fig.\ref{fig:spin}-a) or the hyperfine coherence $\Delta m =0$ (Fig.\ref{fig:spin}-b) and their relaxation at high atomic densities is dominated by spin–exchange collisions.
However, Zeeman coherence with $\Delta m =1$ associated with the spin orientation moment (Fig.\ref{fig:spin}-c) has been shown to be unaffected by spin-exchange collisions at low magnetic fields \cite{happer1973spin,shaham2020quantum}. 
Using a decoherence-free subspace of Zeeman coherences, coherent light storage for 1 second in a hot vapor of caesium has been demonstrated using a mapping of the light field onto the spin orientation \cite{katz2018light}.

An exciting prospect for this work is to use the ultra-stable nuclear spin of noble gases to reach coherence times on the scale of hours \cite{walker1997spin}.
However, the isolation of these states limits the ability to interface with other quantum systems coherently, since they are not optically accessible.
Recent theoretical investigations have explored possible ways to create a quantum interface for noble-gas spins, without limiting their long coherence time, using spin-exchange collisions with alkali-metal atoms \cite{PRXQuantum.3.010305}.
For example, a theoretical scheme for entangling two macroscopic ensembles of noble-gas spins contained in distant cells has been proposed \cite{katz2020long} where the alkali mediators are optically-accessible and couple to the noble-gas spins via coherent spin-exchange collisions \cite{PRXQuantum.3.010305} and the coherent bidirectional coupling between light and noble-gas spins has been demonstrated using vapors of $^{39}$K and $^3$He \cite{katz2021coupling}.

\section{Hot vapors for quantum fluids of light}
\label{QFL}
In this last section, we will discuss an emerging field for hot atomic vapors: paraxial fluids of light.
Due to the mathematical similarities between the equation describing the non-linear propagation of a laser beam in a Kerr medium \cite{2003_boyd} and the non-linear Schrödinger equation describing the temporal evolution of quantum gases with contact interactions, it is possible to draw fruitful analogies between these two systems.
In this section, we will recall the history of this approach, describe the associated mathematical mapping and present recent results and future prospects for this field.

\subsection{General concept}
Within the paraxial and slowly-varying envelope approximations, the propagation equation for a laser inside a centro-symmetric and isotropic non-linear Kerr medium takes the form: 
\begin{equation}
    \label{eq:NLSE}
    i \frac{\partial \mathcal{E}_0}{\partial z} \ (\mathbf{r}_\perp, z ) =  \left[ -\frac{1}{2 n_0 k_0} \boldsymbol{\nabla}_\perp^2 - k_0 \delta n(\mathbf{r}_\perp, z) - \frac{k_0}{2 n_0} \chi^{(3)}(\omega_L) |\mathcal{E}_0 (\mathbf{r}_\perp, z )|^2  \right]\mathcal{E}_0 (\mathbf{r}_\perp, z )~.
\end{equation}
In Eq.~\ref{eq:NLSE}, we have introduced the slowly varying envelope of the electric field $\mathcal{E}_0$ and its wavevector $k_0$.
$n_0$ is the linear refractive index given by $n_0 = \sqrt{1 + \textrm{Re}\chi^{(1)}}$.
The Laplacian term plays the role of a kinetic energy with an effective mass $m^* \propto n_0 k_0$.
The nabla operator $\boldsymbol{\nabla}_\perp$ must be understood as acting in the transverse $\mathbf{r}_\perp=(x,y)$ plane as a consequence of the paraxial approximation.
The second term $k_0 \delta n(\mathbf{r}_\perp, z) \mathcal{E}_0 (\mathbf{r}_\perp, z )$ can be interpreted as an external potential acting on photons in the transverse plane, where $\delta n(\mathbf{r}_\perp, z )$ is a local variation of the linear refractive index.
Finally, in the monochromatic approximation, the non-linear term is proportional to the third-order susceptibility at the laser frequency $\chi^{(3)}(\omega_L)$ times the laser intensity.
In analogy with atomic quantum gases, this term induces an effective photon-photon interaction, depending on the sign of $\chi^{(3)}(\omega_L)$.
A laser beam (with a gaussian profile) propagating through a $\chi^{(3)}$ non-linear medium will either experience self-focusing (for $\chi^{(3)}(\omega_L)>0$) yielding attractive interactions or self-defocusing (for $\chi^{(3)}(\omega_L)<0$) giving rise to repulsive interactions between photons.

\begin{figure}[]
    \centering
	\includegraphics[width=0.6\textwidth]{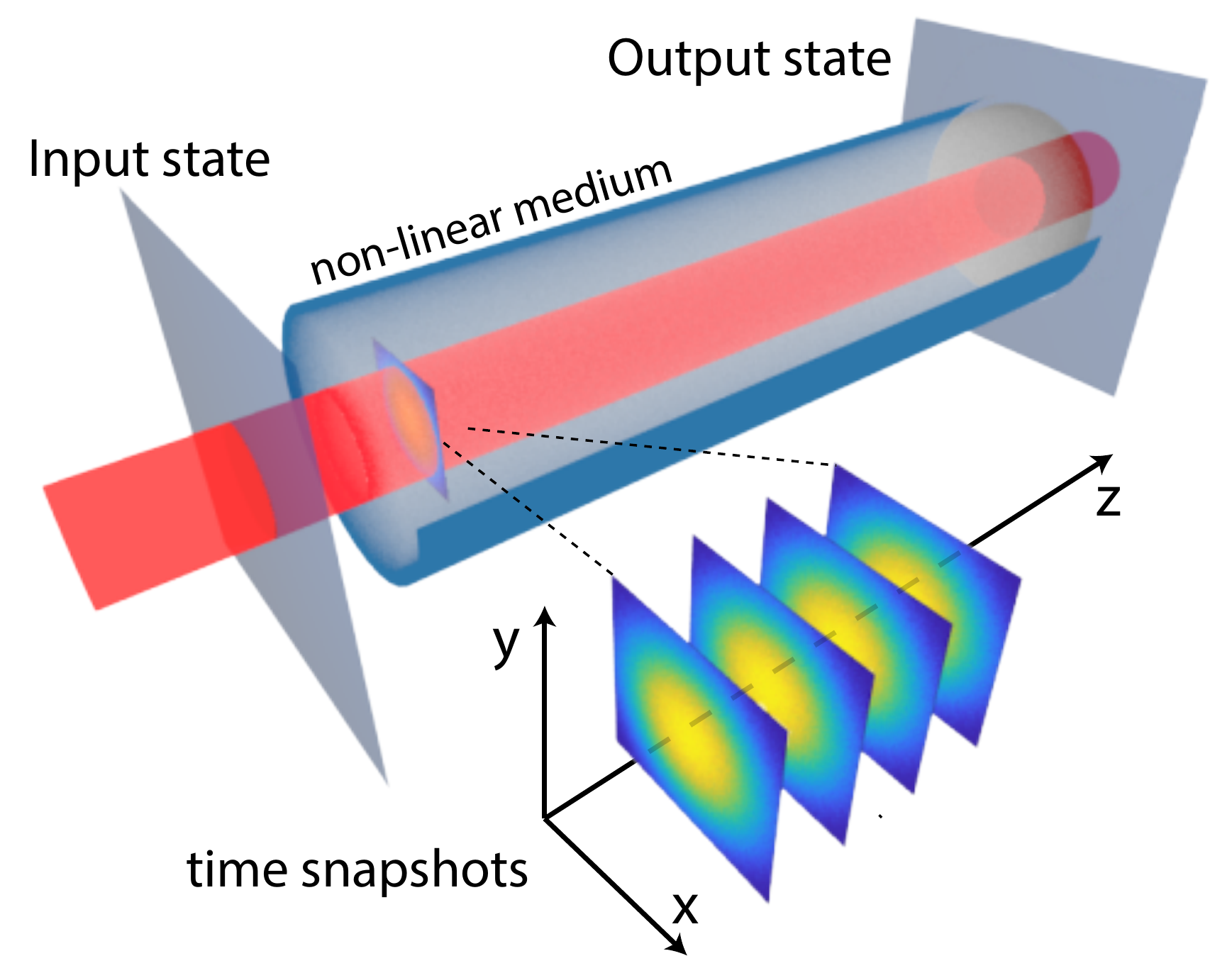}		
   \caption{Space-time mapping in a paraxial fluid of light configuration. Transverse electric field during the propagation of a laser in a non-linear medium is interpreted as time snapshots of a 2D quantum fluid.  The figure is taken from \cite{fontaine2020interferences}}
	\label{fig:manip}
\end{figure}

Bridging the gap between this spatial formulation and the temporal non-linear Schrödinger equation (NLSE) (Eq.~\ref{NLSE-t}), simply requires mapping the $z$ axis into an effective time through the space-time mapping: $z \leftrightarrow \tau = z n_0 /c$, with $c$ the speed of light in vacuum \cite{carusotto2014superfluid}.
Within this picture, the 2D+1 evolution of the electric field along the propagation direction $z$ is a succession of time snapshots illustrated in Figure \ref{fig:manip}.

While this formalism applies to all materials with Kerr non-linearity, including photo-refractive crystals \cite{michel2018superfluid,boughdad2019anisotropic,eloy2021experimental} and thermo-optic liquids \cite{vocke2015experimental,vocke2016role,braidotti2022measurement}, hot atomic vapors are a media of choice since they allow a precise control of the linear and non-linear index of refraction through optical manipulation of the atomic states (optical pumping, coherent effects...).

\subsection{From nonlinear optics to Gross-Pitaevskii equations}
Initially proposed by Ginzburg and Pitaevskii \cite{ginzburg1958sov}, a common approach to describe neutral superflows is to use the NLSE:
\begin{equation}
\label{NLSE-t}
    i\frac{\partial \Psi}{\partial t} = \frac 12 \Delta \Psi - |\Psi|^2 \Psi.
\end{equation}
It is then possible to map the NLSE into a fluid-like equation by using the Madelung transform \cite{madelung1927quantentheorie} ($\Psi=\rho\ \e^{i \phi}$) where the fluid density $\rho=|\Psi|^2$ and the fluid velocity $v=-\nabla \phi$:
\begin{equation}
    \frac{\partial \rho}{\partial t}=\nabla \cdot (\rho\nabla 
\phi) \\ \textrm{ and } \\
\frac{\partial \phi}{\partial t}=-\frac{1}{2\sqrt{\rho}}\nabla^2(\sqrt{\rho})+\frac 12 (\nabla \phi)^2+\rho.
\end{equation}
Several early theoretical works mention the possibility of studying phase singularities (vortices, solitons) \cite{frisch1992transition} and hydrodynamic effects in optics following this approach \cite{MATTAR19811}. 
To complement this review, we provide a practical \texttt{NLSE} solver, written in Python \cite{10.5555/1593511}, that harnesses the power of modern GPU (graphics processing unit) computing capabilities.
The code is open source and fully available in the supplementary material \cite{aladjidi_2022_6918514}. The details of the implementation can be found in the supplementary section \ref{section:supplementary}.

Within this framework, the concept of fluids of light has emerged, experimentally, around microcavity-exciton-polariton systems (see \cite{2013_carusotto} and references therein for a review) with the demonstration of polariton Bose-Einstein condensation \cite{kasprzak2006bose} and polariton superfluidity \cite{2009_amo}.
In these experiments, the non-linear material (semi-conductor quantum wells) are placed within a high finesse cavity.
Strong coupling between cavity photons and quantum well excitons leads to bosonic pseudo-particles (polaritons) following a driven-dissipative dynamics.
To improve the control and tunability of the active medium non-linearity, atomic vapors have been proposed as an interesting candidate for complementary systems to study superfluidity \cite{chiao1999bogoliubov,chiao2000bogoliubov}.
However, only few experiments were conducted at the time, with, unfortunately, no demonstration of superfluidity of light in a hot atomic vapor \cite{mccormick2003transverse,chiao2004effective}.

\subsection{Early experiments with hot atomic vapor cells}

The power of using atomic resonances to enhance nonlinear optics has been recognized shortly after the advent of lasers. This has led to a large number of atomic physics experiments using hot atomic vapor cells, well before the development of laser cooled atoms. As the Doppler broadening is larger than the natural linewidth of the most commonly used transitions, Doppler-free spectroscopy techniques have been implemented to overcome this strong inhomogeneous broadening. However, exploiting the Lorentzian lineshape of the natural linewidth compared to the Gaussian linewidth of Doppler broadening, it is possible for sufficiently large detuning to explore physical phenomena in hot atomic vapors as if the system were at zero temperature. This strong feature has allowed hot atomic vapors to be modeled by simple models. In addition, it is easily possible to reach spatial density combined with large sample lengths, orders of magnitude beyond what can be reached with laser cooled atomic samples. The price to pay to access this regime is the large detuning (typically of the order of several GHz) and the corresponding larger laser power (typically of the order of W) to study nonlinear optics with hot atomic vapors. 
It is obviously impossible to give a full overview of all the nonlinear optics experiments performed over the years. We will thus restrict this short historical overview to some experiments in a single pass propagation geometry that are most relevant to the more recent experiments which were conducted after the connection to the Gross-Pitaevskii equations was demonstrated, renewing interest in these experiments.

Very early single-pass propagation experiments illustrating the resonant nonlinear optical effects with atomic vapors have been reported in \cite{grischkowsky1970,armstrong1972,ashkin1974}. In these experiments self-focusing, self-defocusing and self-trapping were observed in a hot vapor cell of potassium or sodium atoms.  In \cite{grischkowsky1970,armstrong1972,ashkin1974} a cw dye laser was used to excite the atoms close to the D2 line with positive and negative detuning in the GHz range to exploit the nonlinear effects of the anomalous dispersion of the atomic line. In  \cite{happer1977}, an interesting (and yet to be fully exploited) feature has been reported with the observation of long range interactions between opposite polarized laser beams crossing a hot sodium vapor cell excited at positive detuning close to the D1 line. These interactions are based on optically pumped atoms and thus go beyond the simple two-level model used in many subsequent experiments. In another sodium experiment \cite{kaplan1991} with a laser tuned below the D2 line (i.e., with negative detuning), dark soliton stripes in defocusing media have been observed by using suitable initial shapes of the laser, corresponding to different initial conditions of the non-linear propagation equation. This work is an early illustration of the richness of this platform accessed by changing the initial conditions of the equation to be solved. A very important contribution to this field of research has been reported in \cite{kivshar1996}, where vortex solitons created by the instability of dark soliton stripes have been observed. This work can be considered as a pioneering  connection between nonlinear optics and turbulence, a topic actively studied in recent years. Let us also mention the work of \cite{gauthier2002}, where transverse pattern formation was observed in a single pass of a laser through hot sodium vapor cells.

In addition to hot vapor cells containing only alkali atoms, qualitatively new features can arise by adding a buffer gas into the cell. The motion of atoms can be drastically modified from ballistic to diffusive motion. Radiative properties can be controlled using different buffer gases, with e.g. helium or argon allowing for multiple scattering to be maintained \cite{araujo2021} whereas radiative quenching will occur with nitrogen as a buffer gas \cite{lippi1998}. An illustration of novel qualitative effects observed in presence of buffer gas is the so-called optical piston effects \cite{woerdman1984}.

Even though it goes beyond the single-pass configuration, let us also mention that single-mirror feedback schemes, as proposed in \cite{firth1991}, have been implemented in hot atomic vapors  \cite{grynberg1994, lange1994}, with beautiful spatial patterns emerging. This single-mirror feedback scheme has important differences to cavity physics, as there is no imposed transverse mode selection. It will be interesting to investigate a connection between nonlinear optics and nonlinear Schrödinger equations in such more complex settings.

\subsection{Mean field approach: condensation and superfluidity}
Interestingly, experiments with paraxial fluids of light in hot atomic vapors have followed the same path as exciton-polariton systems, with two seminal experiments demonstrating (pre)-condensation and superfluidity.

Bosonic character of the particles is crucial to observe Bose-Einstein condensation (BEC). 
However, it is known that classical waves exhibit a phenomenon of condensation, whose thermo-dynamic properties are analogous to those of the genuine quantum BEC, despite the classical nature of the system \cite{PhysRevLett.95.263901,sun2012observation}.
Indeed for waves traveling in random directions in a nonlinear medium, wave thermalization and condensation can occur. 
The process of two-dimensional thermalization has been observed by propagating a laser beam, that is first rendered spatially incoherent by passing through a diffuser, through a hot rubidium vapor  \cite{vsantic2018nonequilibrium}.
In this experiment, the output of a 1 W fibre laser, tuned below the D2 line of Rb at 780 nm (allowing for defocusing non-linearities), was used to realize a speckle providing a Gaussian distribution of incident wave vectors in the transverse plane.
The light transmitted after nonlinear propagation was analyzed either in density space through direct imaging on a camera or by recording the momentum distribution (in the Fourier plane of a lens).
It was found that the transmitted intensity distribution $P(I)$ strongly deviates from Gaussian statistics, with a maximum of
$P(I)$ that gradually shifts away from $I=0$, as the light propagates.
This deformation reflects a reduction of intensity fluctuations that precedes the establishment of long-range phase coherence that is well established in 2D atomic BEC \cite{PhysRevLett.84.2551}.
This experiment gives an overview of the interest of fluid of light experiments: a relatively simple setup (as usual with hot atomic vapors) and the possibility to study the time evolution of non-equilibrium effects.
While achieving complete thermalization and condensation of random nonlinear waves through nonlinear optical propagation is known to require prohibitively large interaction lengths \cite{Chiocchetta_2016}, in this experiment the phenomenon of pre-condensation far from thermal equilibrium is observed for short propagation lengths.

It is important to understand, how the temporal evolution is monitored in paraxial fluids of light experiments.
As explained in Eq.~\ref{eq:NLSE}, the propagation direction $z$ plays the role of an effective time, but imaging within a non-linear medium is not an option.
Alternatively, one can instead re-scale the effective time by incorporating fluid interaction \cite{abuzarli2021blast} into a new variable $\tau=~{L}/{z_{NL}}$, where  $z_{NL}=\frac{1}{|g| \rho(L)}$ is the non-linear axial length, with the density. $\rho(L)$, defined by the Madelung transform and the interaction coefficient $|g|=\frac{k_0}{2 n_0} |\chi^{(3)}|$.
$L$ is the length of the atomic vapor cell.
After re-scaling the transverse quantities ($\tilde{\textbf{r}}=\textbf{r}/\xi$, $\tilde\nabla_{\perp} = \xi\nabla_{\perp}$) by the transverse healing length $\xi= \sqrt{\frac{z_{NL}}{k}}$, one obtains for $\psi=\frac{\mathcal{E}_0}{\sqrt{\rho(L)}}$:
\begin{equation}
    i\frac{\partial\psi }{\partial \tau}=
    \left(-\frac{1}{2}\tilde\nabla^2_{\perp}+{\mid}\psi{\mid}^2
    \right)\psi.
    \label{GPE_Adim}
\end{equation}

As a direct consequence of Eq. \ref{GPE_Adim}, it is possible to simulate the temporal evolution in these systems by increasing the value of $\tau=~{L}/{z_{NL}}$, i.e by reducing the value of $z_{NL}$. 
Since the non-linear susceptibility in hot atomic vapors can be finely tuned by changing the detuning from resonance or the atomic vapor density \cite{glorieux2018quantum}, paraxial fluids of light in hot atomic vapors are perfectly suited to study the temporal evolution of non-equilibrium phenomena.

The second milestone in establishing paraxial light fluids in hot vapors on a solid experimental basis is the observation of superfluid behavior for light.
In a similar experimental configuration to that described above, a quasi-homogeneous 2D-fluid of light was created by propagating a broad and intense laser beam near atomic resonance into a warm rubidium vapor cell.
A second beam (from the same laser) was overlapped at the cell entrance with a small angle $\theta$.
This beam was narrower and weaker than the fluid beam, in order to create a localized wavepacket of density perturbation.
The wavelength $\Lambda$ of the perturbation is given by $2\pi/k_{\perp}$ where $k_{\perp} = k_{0} \sin \theta$.
The group velocity for a given $k_{\perp}$ was recorded by measuring the transverse distance traveled by the wavepacket after  propagation in the non-linear medium. 
A numerical integration over $k_{\perp}$ of the group velocity directly gave access to the dispersion relation for the weak perturbation on this photon fluid.
The striking result obtained in \cite{fontaine2018observation} is a dispersion relation with a sonic-like behavior ($\propto k_{\perp}$)  at low momenta and particle-like behavior ($\propto k_{\perp}^2$) at high momenta.
Moreover by changing the fluid density, it was possible to demonstrate that the speed of sound (defined as the low momenta proportionality constant) scaled as the square root of the density, which is the exact same behavior as in atomic superfluids \cite{pitaevskii2016bose}. 
More recently, it has also been shown that these sound-like elementary excitations in the photon fluid (i.e. the phonons) could interfere and modify their propagation behavior \cite{fontaine2020interferences}. 
These results open the way to the in-depth study of quantum gas physics and especially out-of-equilibrium dynamics in fluids of light.

\subsection{Emergence of transverse coherence: the Berezinsky-Kosterlitz-Thouless transition}
In the previous section, we described the spectral redistribution of intensity for classical waves propagating in a non-linear medium.
These works have been recently complemented by measurements on the transverse spatial coherence of the fluids of light.
The emergence of a long range coherence is the hallmark of Bose-Einstein condensation in three dimensions. 
In two dimensions, a geometry relevant for fluids of light in the paraxial configuration, the phase transition does not occur at non-zero temperature due to the thermal fluctuations.
However, a quasi long range order is possible (even without interactions) related to the emergence of an extended phase coherence across the system.
Moreover, in the presence of interactions,  the situation gets more complicated with the possibility of a superfluid - Berezinskii–Kosterlitz–Thouless (BKT) transition.

With a light field, transverse coherence can be measured by monitoring the contrast of an interference pattern. 
Experimentally, for an azimuthaly symmetric situation, this is done by splitting a beam in two paths, and inverting one arm with a pair of Dove prisms or a retro-reflector.
After recombination, the contrast at a given distance from the center of rotation gives a direct measurement of the local spatial coherence.
Following a theoretical proposition of  \cite{PhysRevResearch.2.013297}, this procedure has been implemented in a fluid of light experiment, to test the thermalization process in two dimensions \cite{Abuzarlib}.
In this work, the authors added a small tunable fraction of random spatial fluctuations to a fully (spatially) coherent beam using a spatial light modulator with a controllable modulation depth.
This spatial noise is interpreted as a tunable amount of kinetic energy injected in the system initially at zero temperature. 
Since the system is intrinsically out-of-equilibrium, because the thermalization process begins only at the non-linear medium entrance due to the interactions, a temperature cannot be properly defined, but the amount of kinetic energy injected in the system can control a phase transition.

In \cite{Abuzarlib}, the authors report the emergence of a long range coherence in a two-dimensional fluid of light with this approach with two original aspects:
i) the appearance of long range coherence propagates at the speed of sound and therefore is limited by a light cone, ii) the long range coherence disappears when the amount of injected kinetic energy is above a certain threshold, mimicking an out-of-equilibrium counterpart of the BKT transition.
Interestingly, these results are complemented by measurements of the number and correlation properties of free vortices in another non-linear system (photorefractive crystal) \cite{Situ} reinforcing the hypothesis of a BKT-like transition in non-equilibrium optical systems.

These experiments, together with the experiments on non-equilibrium precondensation of classical waves \cite{vsantic2018nonequilibrium} illustrate the potential of hot atomic vapors to contribute to the question of spontaneous emergence of spatial coherence in nonlinear media. Whereas in 2D, in the absence of an external potential, BKT-like equilibrium or non-equilibrium phenomena can be studied, it is also possible (although not yet observed experimentally) to add an external confining potential allowing for Bose-Einstein condensation to be explored.

\subsection{Fluids of light in the presence of an external potential and simulation of non-equilibrium phenomena}

A direct field of application of paraxial fluids of light is the study of non-linear out-of-equilibrium effects such as dispersive shockwaves and turbulence.

In hydrodynamics, but also in plasma physics or ultracold quantum gases, the short time propagation of slowly varying nonlinear pulses can be described discarding the effects of dispersion and dissipation.
This treatment predicts that, due to non-linearity, an initially smooth pulse steepens during its time evolution and reaches a point of gradient catastrophe.
This is the wave-breaking phenomenon, which results in the formation of a shock wave \cite{Zeldovich-Raiser} which eventually acquires a stationary nonlinear oscillating structure \cite{Sagdeev1966}. 

The formation of shockwaves in a hot atomic vapor has been studied in detail. 
The general framework consists of overlapping a wide fluid beam with an intense perturbation (typically the same intensity or larger than the fluid).
Temporal evolution is then recorded by changing the non-linear coefficient as explained previously.
A quantitative description of shockwaves in quasi 1-dimension has been proposed in \cite{bienaime2021quantitative} by taking into account saturation of the non-linearity and dissipation using Whitham modulation theory.
This result has been obtained by also considering non-local interactions that arise when the atomic vapor temperature is greater than 150 $^{\circ}$C \cite{azam2021dissipation}.
Since non-locality also brought novel physics, this configuration led to a double shock-collapse instability: a shock (linked to the gradient catastrophe) for the velocity as well as a ring-shaped collapse singularity for the density.
This instability was explained by the combined effect of the non-local photon-photon interaction and the linear photon losses.
Finally, blast waves have been observed in fluids of light \cite{abuzarli2021blast}.
A blast wave is characterized by an increased pressure and flow resulting from the rapid release of energy from a concentrated source and its particular characteristic is that it is followed by a wind of negative pressure, which induces an attractive force back towards the origin of the shock. 
By comparing 1D and 2D geometry (in the transverse plane) it has been observed that blast waves in optics only occur in 2D, as predicted from the hydrodynamical model \cite{friedlander1946diffraction}.

Turbulent phenomena have been widely explored in atomic superfluids, and paraxial fluids of light could be a complementary platform to study the onset of turbulence instability and short time evolution.
Using initial conditions with a depleted central core combined with a controlled phase difference to a background fluid, spontaneous generation of vortices have been observed in \cite{azam2022}. After an initial development of snake instabilties, vortices formed and could be steered to either move away or closer together. In the later case, annihilation of vortices with radiative losses have been observed. An in situ wavefunction tomography of the vortices with amplitude and phase mapping has been performed using a wave-front sensing camera. 
For fully developped turbulence, no experimental results are known, so far, but two interesting theoretical proposals should be mentioned.
First, by colliding two counter-streaming fluids of light, it has been shown \cite{rodrigues2020turbulence} that the nucleation of a high density of vortex-antivortex pairs, similar to the so-called ultraquantum regime can be observed in a atomic vapor.
Moreover, the absence of a direct energy (Kolmogorov) cascade in the kinetic energy spectrum is demonstrated.
The experimental implementation consists of preparing two fluids (by splitting a near resonance laser in two) and launching these two fluids with opposite transverse relative velocities.
This point is another strong advantage of fluids of light: the density and velocity of the initial fluids can be defined at will by optical manipulation.
For example, here, by using a spatial light modulator (SLM) one can shape the phases of the two fluids and therefore tune the relative velocities.
A similar configuration has been proposed to probe the  Kelvin–Helmholtz instability at the interface between two counter-propagating superflows \cite{giacomelli2021interplay}.
In this case, instead of having the two beams collide directly, the phase profile is engineered to create an interface and generate vortices, which is also realistic with a SLM.\\

In the discussion so far, we have set the potential energy term to zero, which leads to neither trapping nor repulsive forces acting on the fluid.
It is possible however to engineer the light-matter interaction in the hot atomic vapors to control locally the index of refraction and therefore induce a local potential (either repulsive or attractive).
The simplest approach is to use optical pumping to locally modify the atomic density and therefore the index. 
As the propagation direction $z$ is the effective time, a constant defect in time requires a constant modification of the refractive index along $z$. 
This can be done by using non-diffracting Bessel beams and compensating for the absorption \cite{fontaine2019attenuation}.
A more complex approach relies on coherent effects in the atomic medium.
As described in the previous section, EIT will induce a strong modification of the refractive index of an atomic medium.
This idea has been exploited to create optical lattices for fluids of light.
For example, photonic graphene has been realized in hot atomic vapors \cite{zhang2019particlelike}.
By interfering 3 laser beams inside a rubidium vapor cell, Zhang et. al, created a honeycomb lattice using EIT.
By changing the one photon detuning they were able to modify the susceptibility of the lattice and therefore tune the amplitude of the potential and the tunneling probability between the lattice sites.
By injecting optical vortices in linear wave packets, using Laguerre-Gauss beams, they showed that vortex trajectories obey particle-like behaviour, in particular they reported the effect of the Magnus force and the mutual interaction of two vortices. 
In a similar configuration, edge states have been observed in photonic graphene \cite{zhang2020observation}.
It opens the way for experimental exploration of nonlinear topological photonics in hot atomic vapors which has the advantage of being a fully reconfigurable platform.

One more important ingredient can be added to this configuration.
With manipulation of atomic coherences, it is not only possible to tune the real part of the susceptibility (the refractive index) but also the imaginary part (linear loss or gain).
This has particular interest since it allows for designing non-Hermitian Hamilitonians with parity-time (PT) symmetry.
In the fluid of light analogy, the real and imaginary parts of the refractive index correspond respectively to spatially symmetric real and anti-symmetric imaginary parts of a complex PT-symmetric potential.

A scheme for creating PT symmetry based on atomic coherence is suggested in a three-level rubidium vapor mixture of $^{85}$Rb and $^{87}$Rb using interference between two Raman resonances \cite{PhysRevLett.110.083604}.
Experimentally, a PT-symmetric Hamiltonian in a optical lattice with periodical gain and loss profiles has been realized using coherence manipulation with 4 atomic levels \cite{zhang2016observation}. 

A remarkable alternative approach makes use of the atomic motion in a hot vapor to engineer an anti-PT-symmetric system \cite{peng2016anti}.
The coupling between two spatially separated fields is mediated through coherent mixing of spin waves created in two parallel optical channels, thanks to atomic diffusion.
Since this configuration relies on the long-lived ground-state coherence the spin-wave diffusion enables surprising linear and nonlinear interactions between the two parallel fields
For an extensive review of PT-symmetry in optics, the interested reader could refer to \cite{Zhang_2018}.

\subsection{Beyond the mean field approximation}
The next frontier in paraxial fluids of light is to go beyond the mean field approximation used to derive Eq. \ref{eq:NLSE}.
Originating from statistical physics, this terminology is widely used for ultra-cold atomic quantum gases and consists in averaging the interaction effects exerted on each single particle by the other ones with a local density term.
For revealing the many-body quantum effects beyond this approximation a quantum description of the field is required.
Hot atomic vapor is an excellent candidate to pursue this goal, since highly non-classical and entangled states have been produced in this system, as described earlier in this review.
From an optics perspective, going beyond the mean field approximation is similar to going from non-linear optics to quantum optics with paraxial fluids of light.
Theoretically, a complete quantum description has been proposed in \cite{larre2015propagation}.
This work uses quantized operators for the fields and introduces a third effective spatial dimension to the problem which is proportional to the physical time.
The propagation direction plays the role of an effective time, and the physical time plays the role of a third spatial dimension (assumed to be frozen in the  case of a purely monochromatic laser).
A direct consequence of going beyond the mean field regime is quantum depletion. 
Even at zero temperature, a weakly interacting quantum fluid at equilibrium will have a fraction of its population distributed away from the condensate (i.e. the zero momentum part).
This depleted part is a direct consequence of the quantization (and the bosonic commutation rules) of the elementary excitations in the quantum fluids.
For fluids of light, this can be understood as a four-wave-mixing process: taking two excitations out of a paraxial fluid  $k_{\perp} = 0$ and redistributing in two modes at $k_{\perp} \neq 0$.
Since, we are considering only 2-body contact-like interactions, all the collisions can be interpreted as four-wave-mixing processes (in the transverse plane).
A direct measurement of the depleted fraction is a notably hard experimental task since it requires isolating the weak $k \neq 0$ contribution (scaling as $k^{-4}$) over a large background at $k=0$; and has only been achieved recently in ultra-cold atoms \cite{lopes2017quantum,tenart2021observation}.
However, it is possible to gain insight on the quantum depletion, by measuring the static structure factor $S(k)$ of a fluid.
In optics, the static structure factor translates to the spatial noise spectrum.
It is the spatial analogue of the usual quantum optics temporal noise spectrum (normalized to the shot noise) described in the first section of this review.
In \cite{piekarski2021measurement}, a static structure factor below 1 has been measured at low $k_{\perp}$.
This is a direct indication that quantum effects are indeed present in fluids of light propagating in hot atomic vapors.

Finally, an important aspect of paraxial fluids of light, is the presence of two quenches of interactions, at the entrance to the  medium and at the exit.
It is known that quenching, i.e. changing non-adiabatically, one parameter in a physical system will create spontaneous excitations in the density that actually warms the system.
These excitations also remove part of the population away from the ground state at $k_{\perp}=0$ which depends on the \textit{non-adiabaticity} of the quench and can be distinguished from depletion (which is an equilibrium effect).
A recent experiment has shown evidence of these spontaneous excitations \cite{steinhauer2021analogue}. 
A pulsed laser (100 ns pulses) is sent through a hot atomic vapor of rubidium near resonance and the output of the medium is imaged on a camera.
After 1000 realizations, the average density is calculated and subtracted from each of the images to extract the spatial fluctuations.
A spatial Fourier transform of the fluctuations gives access to the spatial noise spectrum $S(k)$.
The same procedure was iterated for other image planes (after the medium) to record the evolution of the fluctuations generated at the second quench (the cell output) and a clear signature of the presence of excitations spontaneously generated by the quench is visible.
This work opens the way to study non-equilibrium many-body physics with light, with more sophisticated detection schemes such as spatio-temporal homodyne measurements.

 \section{Conclusions}

 In this review paper we have addressed several aspects of nonlinear optics with hot atomic vapors, mainly squeezing, quantum memories, and quantum fluids of light.
 A long-term goal of the investigation non-classical light is to actually take advantage of the noise or correlation properties in measurements in a system that can potentially be moved out of the lab \cite{Lawrie}.  A number of specific demonstrations of the use of squeezed light generated in hot vapors have been made, including in plasmonic sensors \cite{Dowran} \cite{Lee}, cantilever sensors in atomic force microscopy \cite{Pooser2}, interferometry \cite{Anderson} \cite{Jing}\cite{Prajapati}, and especially magnetometry \cite{Horrom} \cite{Otterstrom}.
The enhancement of two-photon absorption using twin-beam sources of correlated light for the possible improvement of the sensitivity of two-photon microscopy is also a current topic of interest \cite{Tian}.

From the early days of quantum memories to the most recent works, hot atomic vapors have always been an exciting platform to explore original approaches and test novel protocols. The recent results of \cite{katz2021coupling} open the way to hour-long quantum memories in hot atomic vapors. 
Even though only one quadrature is coupled coherently to the noble-gas spins so far, it is possible to achieve a complete mapping at high noble-gas pressure,  by  sending  the  light  beam through the cell twice \cite{muschik2006efficient}.
Overall, hot atomic vapors, by their experimental simplicity and availability, will certainly play a major role in the coming years for the implementation of long distance quantum networks and the quantum internet \cite{wehner2018quantum}.

Paraxial fluids of light (commonly called quantum fluids of light in recent years) in hot atomic vapors have emerged as a exciting platform for studying quantum gases, with the advantages of optical setups (compactness, precision of the detection methods, wavefront-shaping) and the possibility offered by these experimental configurations to study non-equilibrium effects.
Moreover, the fluid of light approach brings a different understanding of non-linear and quantum optics phenomena and it would be interesting to reconsider previous work in quantum optics such as diffraction elimination in hot atomic vapors \cite{firstenberg2009elimination,firstenberg2009elimination2}, Levy flights \cite{mercadier2009levy} or weak coherent back scattering \cite{cherroret2019robust} and weak localization \cite{cherroret2021weak} in this quantum fluid framework to see if an original perspective on these topics could emerge.
 
 \section{Acknowledgments}
 We thank the many members of our research teams over the last years for their fruitful and stimulating interactions. 
 We acknowledge financial support for Air Force Office of Scientific Research (FA9550- 16-1-0423), the FET Flagship project PhoQuS (agreement No 820392), the Agence Nationale de la Recherche (grants ANR-21-CE47-0009 Quantum-SOPHA, RAN22767 R767 FOLIO), and  the Region Île-de-France in the framework of DIM SIRTEQ.
Q. G. is member of the Institut Universitaire de France (IUF). 

\clearpage

\section{Supplementary information : \texttt{NLSE} solver}
\label{section:supplementary}
 The solver implements the spectral split-step method for solving the non-linear Schrödringer equation. It aims at providing a simple and generic framework to solve most of the Gross-Pitaevskii type equations.  
\paragraph{Code availability} The code is fully documented and readily available at the following link \url{https://doi.org/10.5281/zenodo.6918513}. It is under an open source MIT License.

\paragraph{Features} The solver consists of a single file \texttt{nlse.py} that consists of a single solver class \texttt{NLSE} accompanied with several helper functions. The \texttt{NLSE} class contains all of the relevant physical constants of the NLSE. It assumes the situation of light propagation along an axis $z$ inside a non-linear medium:
\begin{equation}
    \frac{\partial\mathcal{E}}{\partial z} = \Big(-\frac{1}{2k_0}\nabla^2_\perp + \frac{k_0}{2} V -\frac{k_0}{0} n_2 c \epsilon_0 |\mathcal{E}|^2 -i\alpha\Big)\mathcal{E},
\end{equation}
where $\mathcal{E}$ is the electric field in V/m.
Hence, the physical attributes of the class are the following:
\begin{itemize}
    \item $n_2$ the non-linear coefficient in $m^2/W$
    \item $V$ is a dimensionless potential (a local change in the linear index of refraction)
    \item $\lambda$ the wavelength such that the wavenumber $k_0=\frac{2\pi}{\lambda}$
    \item $\alpha$ is the absorption coefficient in $m^{-1}$
    \item $w_0$ the beam waist
    \item $L$ the length of propagation in the non-linear medium.
\end{itemize}
The naming of the attributes is related to configuration described above, however to allow for an easy modification to fit various physical scenarios, the propagation function \texttt{nl\_prop} assume a simplified evolution equation for an arbitrary field $f$ and evolution variable $x$:
\begin{equation}
    \frac{\partial f}{\partial x} = (-\nabla^2 f + V + g |f|^2 -i\alpha)f.
\end{equation}
The main class method is the \texttt{out\_field} method that propagates a given field $E_0$ over a distance $z$, and can plot the final field using the \texttt{plot} option.

\paragraph{Dependencies}  As the solver uses fast fourier transforms (FFT), it works best when employed on a GPU equipped machine, but the code can also work on a regular CPU, albeit more slowly. 

\section{References}

\bibliographystyle{unsrt}
\bibliography{biblio} 

\end{document}